\documentclass[conference]{IEEEtran}
\usepackage{cite}
\usepackage[dvips]{graphicx}
\usepackage[T1]{fontenc}
\usepackage{dcolumn}
\usepackage{bm}
\usepackage{color}
\usepackage{amssymb}
\newcommand{\figjov}[1]{\includegraphics[width=0.45\textwidth]{#1.EPS}}

\newcommand{\Rap}{\mathop{R\mathrm{_{AP}}}}
\newcommand{\Rp}{\mathop{R\mathrm{_{P}}}}

\begin{document}

\title{Dielectric breakdown in underoxidized magnetic tunnel junctions: Dependence on oxidation time and area}

\author{\authorblockN{J.~Ventura}
\authorblockA{IFIMUP and\\Physics Department\\
University of Porto\\
Porto, Portugal\\
Email: joventur@fc.up.pt} \and
\authorblockN{R. Ferreira}
\authorblockA{INESC-MN and\\Physics Department\\
IST\\
Lisbon, Portugal\\
Email: rferreira@inesc-mn.pt} \and
\authorblockN{J. B. Sousa}
\authorblockA{IFIMUP and\\Physics Department\\
University of Porto\\
Porto, Portugal\\
Email: jbsousa@fc.up.pt} \and
\authorblockN{P. P. Freitas}
\authorblockA{INESC-MN and\\Physics Department\\
IST\\
Lisbon, Portugal\\
Email: pfreitas@inesc-mn.pt}}

\maketitle

\begin{abstract}
Magnetic tunnel junctions (MTJs) with partially oxidized 9 \AA~
AlO$_x$-barriers were recently shown to have the necessary
characteristics to be used as magnetoresistive sensors in
high-density storage devices. Here we study dielectric breakdown in
such underoxidized magnetic tunnel junctions, focusing on its
dependence on tunnel junction area and oxidation time. A clear
relation between breakdown mechanism and junction area is observed
for the MTJs with the highest studied oxidation time: samples with
large areas fail usually due to extrinsic causes (characterized by a
smooth resistance decrease at dielectric breakdown). Small area
junctions fail mainly through an intrinsic mechanism (sharp
resistance decrease at breakdown). However, this dependence changes
for lower oxidation times, with extrinsic breakdown becoming
dominant. In fact, in the extremely underoxidized magnetic tunnel
junctions, failure is exclusively related with extrinsic causes,
independently of MTJ-area. These results are related with the
presence of defects in the barrier (weak spots that lead to
intrinsic breakdown) and of metallic unoxidized Al nanoconstrictions
(leading to extrinsic breakdown).
\end{abstract}
\maketitle


Magnetic tunnel junctions (MTJs) consisting of two ferromagnetic
(pinned and free) layers separated by an insulating barrier are the
new generation of magnetoresistive sensors in high-density storage
devices currently approaching 200--400 Gbit/in$^2$
\cite{TMR_ReadHeads}. For MTJs to be implemented in hard drives,
they must have low resistance-area product \mbox{(R$\times$A<1
$\Omega\mu$m$^2$)} and reasonable tunnel magnetoresistance
(TMR>20\%) \cite{Sensors_Parameters}. Such goals can be obtained in
ultrathin tunnel junctions (insulating barrier thickness
$t\sim$5--6~\AA). However, this approach may lead to the presence of
pinholes across the insulating barrier (regions of direct contact
between the pinned and free layers having Ohmic resistance),
resulting in an undesirably large interlayer coupling field. Similar
TJ-characteristics (high TMR, low R$\times$A) were recently obtained
by only partially oxidizing thicker \mbox{(9 \AA)} AlO$_x$-barriers
\cite{Ricardo_LowR_MTJs}. Although decreasing oxidation time lead to
the decrease of both TMR and R$\times$A, fairly high TMR
($\sim20\%$; R$\times$A$\sim2$--$5~\Omega\mu$m$^2$) was still
observed even when the oxidation was performed with the shutter
closed.

Dielectric breakdown (DB) is the main reliability concern in MTJs
\cite{Oepts_Analysis,Oliver2,Oliver3,Breakdown_Rao_thickness,DB_Kim,DB_Scmalhorst}
where it can occur through two distinct mechanisms
\cite{Oliver2,DB_Kim}. \emph{Intrinsic} breakdown occurs in MTJs
with well formed oxide layers due to the action of the applied
electrical field. This leads to an abrupt decrease of the
TJ-electrical resistance (and TMR) as a consequence of the formation
of microscopic ohmic shorts in the barrier
\cite{Oepts_Analysis,DB_gate_oxides}. On the other hand,
\emph{extrinsic} breakdown is related with the growth of existing
pinholes in the tunnel barrier due to localized heating caused by
high electrical current densities flowing along such pinholes
\cite{Oliver3}. In experiments, this mechanism is characterized by a
gradual variation of the electrical resistance at the onset of
breakdown. Because tunnel junctions with thicker barriers have a
lower concentration of pinholes, they fail intrinsically more often
than thinner ones \cite{DB_Kim}.

Here we study dielectric breakdown in underoxidized
CoFeB/AlO$_x$/CoFeB MTJs. Breakdown in these samples occurs at
localized spots of the barrier due to the extrinsic or intrinsic
mechanisms, depending on both TJ-area and oxidation time. For MTJs
with the largest studied oxidation time failure occurs more often
due to extrinsic than intrinsic causes in junctions with large areas
($A\sim10\mu$m$^2$). On the other hand, MTJs with small areas
($A\sim\mu$m$^2$) are observed to fail mainly by an intrinsic
mechanism. Furthermore, with decreasing oxidation time, extrinsic
breakdown becomes the dominant failure mechanism, independently of
MTJ-area. We further observe that extrinsic breakdown is usually
preceded by intense Joule heating caused by large current densities
flowing across unoxidized, metallic Al.

We studied several series of ion beam deposited magnetic tunnel
junctions with different oxidation times \cite{Ricardo_LowR_MTJs}.
The structure of the tunnel junctions studied was: glass/Al (70
\AA)/Ta (90 \AA)/NiFe (70 \AA)/CoFeB (50 \AA)/AlO$_x$ (9 \AA)/X/MnIr
(250 \AA)/Ta (90 \AA), where $X$ is either an amorphous CoFeB (40
\AA) single layer \cite{CoFeB_Susana} or a CoFeB (40 \AA)/Ru (6
\AA)/CoFeB (40 \AA) synthetic antiferromagnet (SAF) structure
\cite{MnPt_Freitas3}. The junctions were patterned to a rectangular
shape with areas between \mbox{$1\times1$ $\mu$m$^{2}$} and
\mbox{$3\times8$ $\mu$m$^{2}$}. The AlO$_x$ barrier was formed by a
remote Ar/O$_2$ plasma (110 W RF in a 20 cm diameter assist ion gun)
\cite{Ricardo_LowR_MTJs}. Ions drift to the chamber due to pressure
gradient only. The oxidation is divided into three stages with total
oxidation time (t$_1$)+(t$_2$)+t$_3$. During the first two stages
(t$_1$)+(t$_2$) the sample is protected by a shutter preventing most
of the oxygen from reaching the sample. The plasma O$_2$ content is
progressively increased: in the first stage the plasma is created
with 4 sccm (Ar)+20 sccm (O$_2$) at a pressure $P=6.5\times10^{-5}$
Torr and in the next two stages one has 4 sccm (Ar)+40 sccm (O$_2$)
at $1.4\times10^{-4}$ Torr. The oxidation time of the studied series
of MTJs ranged from
(25$^{\prime\prime}$)+(00$^{\prime\prime}$)+0$^{\prime\prime}$ to
(30$^{\prime\prime}$)+(30$^{\prime\prime}$)+5$^{\prime\prime}$.

The dependence of the tunnel magnetoresistance on the applied
electrical current TMR($I$) and current-voltage V($I$)
characteristics were simultaneously measured with an automated wafer
probe station. Measurements were performed as follows
\cite{JOV_JAP_twolevelfluc}: under the electrical current $I$ one
measures the resulting voltage drop in the parallel (V$_P$) and
antiparallel (V$_{AP}$) states. The same procedure is performed for
$-I$. The current magnitude is then increased and the above set of
measurements repeated.

\begin{figure}
\begin{center}
\figjov{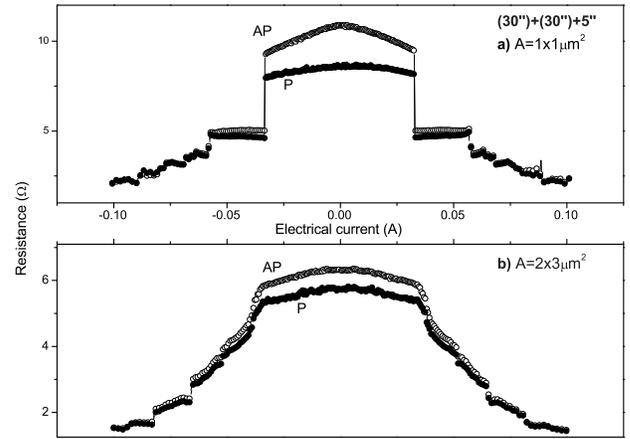}
\end{center}
\caption{Electrical resistance versus applied bias current of a MTJs
with (30$^{\prime\prime}$)+(30$^{\prime\prime}$)+5$^{\prime\prime}$
oxidation time: (a) \mbox{$A=1\times1~\mu$m$^2$} and (b)
\mbox{$A=2\times3~\mu$m$^2$}, for both parallel (P) and antiparallel
(AP) states. Notice the abrupt (smooth) R decrease observed at the
intrinsic (extrinsic) breakdown of the barrier occurring for
\mbox{$A=1\times1~\mu$m$^2$}
(\mbox{$A=2\times3~\mu$m$^2$}).}\label{fig:MR(I)_TJ1001_Int}
\end{figure}

In Fig. \ref{fig:MR(I)_TJ1001_Int}(a) we observe the electrical
resistance versus applied bias current of a CoFe/AlO$_x$/CoFeB/MnIr
MTJ with
\mbox{(30$^{\prime\prime}$)+(30$^{\prime\prime}$)+5$^{\prime\prime}$}
oxidation time and \mbox{$A=1\times1~\mu$m$^2$}. Increasing the
magnitude of the applied electrical current leads to a sudden, sharp
and irreversible R-decrease at \mbox{$|I|\approx32$~mA}. This abrupt
decrease is associated with the \emph{intrinsic} breakdown of the
studied sample \cite{Oliver2}, through the formation of a pinhole in
the barrier. As $I$ is further increased, a new breakdown event is
seen (at \mbox{$|I|\approx55$~mA}), associated with the formation of
a new pinhole.

Tunnel junctions of the same series but with larger area show a
fairly different behavior when DB occurs. Figure
\ref{fig:MR(I)_TJ1001_Int}(b) displays the obtained results for a
MTJ with \mbox{$A=2\times3~\mu$m$^2$}. In this case breakdown (at
\mbox{$|I|\approx35$ mA}), leads to a slight and gradual R (and TMR;
not shown) decrease. This behavior is related with defect-driven
extrinsic breakdown of the barrier reflecting the growth of existing
pinholes \cite{Oliver2,Oliver3} likely through thermally assisted
electromigration of metallic ions from the electrodes into the
barrier \cite{CIS_JOV_IEEE_TN,CIS_JOV_PRB}. Three more breakdown
events are visible at higher currents (\mbox{$|I|\approx50$ mA},
\mbox{$\approx65$ mA} and \mbox{$\approx80$ mA}), further reducing R
and bringing TMR to zero. The new breakdown points are characterized
by small but sharp R-decreases, followed by gradual ones and are
here associated with the formation of new pinholes in the barrier
\cite{DB_Oepts_JMMM} and subsequent current-induced pinhole growth.

Breakdown in these underoxidized MTJs likely occurs at localized
weak-spots of the barrier, where a large concentration of defects
(oxygen vacancies or nanoconstrictions of metallic Al due to the
underoxidation of the barrier) exists. With increasing MTJ-area,
such defects are more expected to appear, leading to the observed
change in the breakdown mechanism from intrinsic to extrinsic. In
fact, for the MTJs with
(30$^{\prime\prime}$)+(30$^{\prime\prime}$)+5$^{\prime\prime}$
oxidation time, we observed that 63\% of those with small area
failed by intrinsic DB, while this number decreased to 20\% for the
large area MTJs.

\begin{figure}
\begin{center}
\figjov{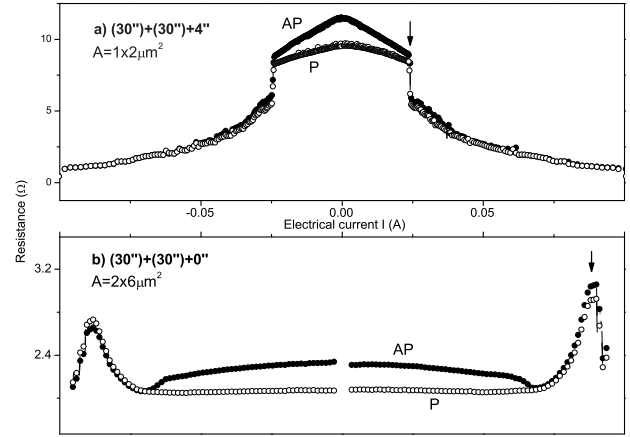}
\end{center}
\caption{Parallel and antiparallel electrical resistance versus
applied current of MTJs with (a)
\mbox{(30$^{\prime\prime}$)+(30$^{\prime\prime}$)+4$^{\prime\prime}$}
and (b) \mbox{(30$^{\prime\prime}$)+(30$^{\prime\prime}$)} oxidation
times.}\label{fig:MR(I)_TJ1022eTJ1018}
\end{figure}

Decreasing the MTJ-oxidation time leads to increased probability of
failure through an extrinsic DB-mechanism for both small and large
area junctions, even though intrinsic breakdown is still observed
[as displayed in Fig. \ref{fig:MR(I)_TJ1022eTJ1018}(a) for a MTJ
with (30$^{\prime\prime}$)+(30$^{\prime\prime}$)+4$^{\prime\prime}$
oxidation time]. Figure \ref{fig:MR(I)_TJ1022eTJ1018}(b) shows
electrical resistance versus bias current for a sample with
(30$^{\prime\prime}$)+(30$^{\prime\prime}$)+0$^{\prime\prime}$
oxidation time, where extrinsic breakdown occurs (see arrow). Notice
the fairly large R-increase just before TJ-dielectric breakdown,
associated with heating due to large current densities flowing
through metallic constrictions across the barrier (see below).

\begin{figure}
\begin{center}
\figjov{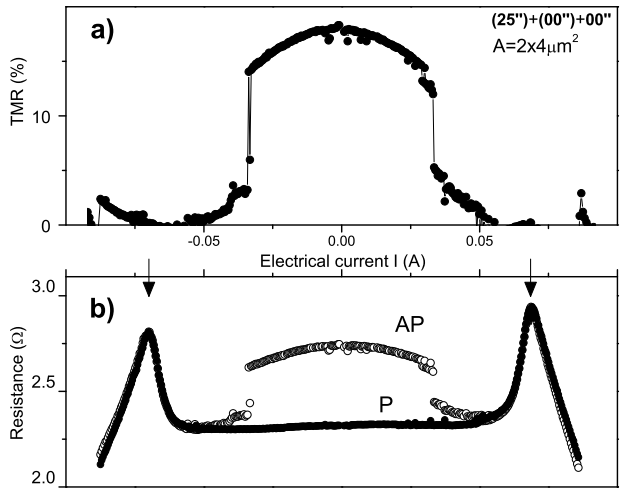}
\end{center}
  \caption{(a) Tunnel magnetoresistance versus bias current [TMR($I$)] and (b) corresponding
  $R(I)$ curves of a MTJ with (25$^{\prime\prime}$) oxidation time.}
  \label{fig_I_V}
\end{figure}

Tunnel junctions with extremely small oxidation time
[\mbox{(25$^{\prime\prime}$)+(00$^{\prime\prime}$)+00$^{\prime\prime}$}]
all fail due to extrinsic reasons after large heating effects.
Figure \ref{fig_I_V} displays TMR(I) and V(I) characteristics for a
MTJ with \mbox{$A=2\times4~\mu$m$^2$} giving \mbox{TMR$\sim15$\%}.
At \mbox{$|I|\approx30$ mA}, TMR(I) sharply decreases. Corresponding
V(I) characteristics for the parallel and antiparallel magnetic
states (not shown) display a quasi-linear behavior also up to
\mbox{$|I|\approx30$ mA}. Fitting our data to Simmons' model gives
the barrier thickness \mbox{$t\approx7~$\AA} and height
\mbox{$\varphi\approx0.5~$eV}. The small values obtained indicate
that only part of the initially deposited Al layer \mbox{(9 \AA)}
was oxidized  and that the oxidized Al is likely not stequiometric.
The observed decrease of TMR is due to the abrupt drop of the
electrical resistance of the antiparallel state ($\Rap$). However,
the electrical resistance of the parallel state ($\Rp$) remains
constant [Fig. \ref{fig_I_V}(b)]. Thus, this decrease is not related
with junction breakdown but is likely due to heating (leading to the
loss of exchange bias) or spin-torque driven instability of the
antiparallel state (due to the high current densities flowing
through unoxidized Al; see below) \cite{CIMS_pillars_3}. At moderate
applied electrical currents, we observe a significant R-increase
[Fig. \ref{fig_I_V}(b)], a behavior that we associate with Joule
heating, due to high current densities flowing through metallic
unoxidized Al nanoconstrictions. The existence of such metallic
paths was indeed confirmed by measurements of the temperature
dependence of the electrical resistance of MTJs of the same series
(equal oxidation time) \cite{JOV_JAP_twolevelfluc}. Only at higher
electrical currents [see arrows in Fig. \ref{fig_I_V}(b)] is the
electrical resistance seen to decrease, due to the extrinsic
breakdown of the barrier. We conclude that nanoconstrictions of
unoxidized Al are the features behind extrinsic dielectric breakdown
in underoxidized MTJs.

Transport in these tunnel junctions then arises from two channels
acting in parallel: one consisting of metallic paths of unoxidized
Al (with resistance R$_m$) and another of tunneling through the
oxidized AlO$_x$ (with resistance R$_t$). We can estimate the area
of the metallic conduction channels assuming that their electrical
resistance is given by the Maxwell formulation
$R_{m}=\frac{\rho}{2a}$ [$\rho$ the electrical resistivity of the
constriction (assumed $\approx10~\mu\Omega$cm) and $a$ the radius of
the metallic channel]. Considering \mbox{$R_m\ll R_t$}, we obtain
for $R\approx2.4~\Omega$ [Fig. \ref{fig_I_V}(b)],
\mbox{$a\approx200$~\AA}, which corresponds to $\approx0.004$\% of
the total TJ-area. This is the upper limit of $a$ since the
tunnel-resistance arising from the oxidized part of the junction
should also be included. The extremely small area of the unoxidized
Al leads to high local current densities and thus to the observed
heating \cite{JOV_Heating}.

In summary, dielectric breakdown in underoxidized MTJs occurs at
localized spots in the barrier, likely where a large concentration
of defects exists. We observed a clear dependency of the breakdown
process on the MTJ area: Failure in MTJs with large areas is usually
of an extrinsic nature, while small area junctions fail mainly by an
intrinsic mechanism. Nevertheless, with decreasing oxidation time,
the extrinsic breakdown becomes dominant, independently of the
MTJ-area. Extrinsic breakdown is preceded by Joule heating caused by
large current densities flowing across unoxidized, metallic Al
nanoconstrictions whose size was estimated.

\section*{Acknowledgment}
Work supported in part by POCTI/CTM/59318/2004, IST-2001-37334 NEXT
MRAM and POCTI/CTM/36489/2000 projects. J. Ventura, and R. Ferreira
are thankful for FCT grants (SFRH/BPD/21634/2005 and
SFRH/BD/6501/2001).

\bibliography{Biblio}

\end{document}